\documentstyle[12pt,aaspp4,psfig]{article}
\newcommand{\cosec}{{\rm \;cosec\; }} 
\newcommand{\etal}{{ et al.\,}}
\newcommand{\ajp}{Australian J. Phys.} 
\newcommand{\ajps}{Australian J. Phys. Astrop. Suppl.}
\newcommand{\pasa}{Publ. Astr. Soc. Australia}
\begin{document}
\title{THE MOLONGLO GALACTIC PLANE SURVEY. \\ I. OVERVIEW AND IMAGES}
\author{A. J. Green, L. E. Cram, M. I. Large \and Taisheng Ye}
\affil{School of Physics, University of Sydney 2006, Australia \\ 
  A.Green@physics.usyd.edu.au, L.Cram@physics.usyd.edu.au,
  M.Large@physics.usyd.edu.au, tye@atnf.csiro.au}

\slugcomment{Submitted to Astrophysical Journal}

\begin{abstract}
  The first epoch Molonglo Galactic Plane Survey (MGPS1) is a radio
  continuum survey made using the Molonglo Observatory Synthesis Telescope
  (MOST) at 843 MHz with a resolution of $43'' \times 43'' \cosec
  |\delta|$.  The region surveyed is $245^\circ \leq l \leq 355^\circ, |b|
  \leq 1.5^\circ$. The thirteen $9^\circ \times 3^\circ$ mosaic images
  presented here are the superposition of over 450 complete synthesis
  observations, each taking 12 h and covering a field of $70' \times 70'
  \cosec |\delta|$.  The root-mean-square sensitivity over much of the
  mosaiced survey is 1--2 mJy beam$^{-1}$ ($1 \sigma$), and the positional
  accuracy is $ \approx 1'' \times 1'' \cosec |\delta|$ for sources
  brighter than 20 mJy.  The dynamic range is no better than 250:1, and
  this also constrains the sensitivity in some parts of the images.  The
  survey area of 330 deg$^2$ contains well over $1.2 \times 10^4$
  unresolved or barely resolved objects, almost all of which are
  extra-galactic sources lying in the Zone of Avoidance.  In addition a
  significant fraction of this area is covered by extended, diffuse
  emission associated with thermal complexes, discrete H II regions,
  supernova remnants, and other structures in the Galactic interstellar
  medium.
\end{abstract}

\keywords{Galaxy: structure --- radio continuum: ISM --- surveys}

\section{Introduction}

The area of sky close to the Galactic equator produces radio emission on a
wide range of angular scales. To unravel the nature of the complex radio
structures seen in the Plane, a high resolution survey is essential. The
Molonglo Observatory Synthesis Telescope (MOST) is well suited to this
task, producing large images ($70'\times 70' \cosec |\delta|$) with
sub-arcminute angular resolution and a sensitivity (1$\sigma$) of 1--2 mJy
beam$^{-1}$.  Accordingly, the southern Galactic Plane has been surveyed
with the MOST in the radio continuum at 843 MHz, covering the area
$245^\circ\leq l \leq 355^\circ$, $|b| \leq1.5^\circ$, with a
resolution of $43''\times43'' \cosec |\delta|$. A report on the early
progress of the survey was published by Whiteoak \etal\markcite{Whi:89}
(1989).  The final data set is a compilation of more than 450 overlapping
fields which have been assembled as a series of $3^\circ\times3^\circ$
mosaics under the title of the First Epoch Molonglo Galactic Plane Survey
(MGPS1). The mosaics are presented here in Galactic co-ordinates, in
panels covering $9^\circ\times3^\circ$ for ease of viewing.

No previous large-scale radio survey of the southern Galaxy has superior
angular resolution or sensitivity. MGPS1 extends and complements other
radio surveys such as the 2.3 GHz Parkes survey (Duncan
\etal\markcite{Dun:95} 1995) with $10'$ resolution, the HI galaxy search
along the Galactic Plane with a resolution of 15$'$ (Staveley-Smith
\etal\markcite{Sta:98} 1998), several Parkes surveys at 5 GHz (Griffith,
Wright \etal\markcite{Gri:93} 1993; Haynes, Caswell \& Simons
\markcite{Hay:78}1978; Goss \& Shaver \markcite{Gos:70}1970), and the 408
MHz Mills Cross surveys (Shaver \& Goss \markcite{Sha:70} 1970; Green
\markcite{Gre:74}1974).  The 1.4 GHz VLA\footnote{the Very Large Array of
  the National Radio Astronomy Observatory} Sky Survey (NVSS) of Condon
\etal\markcite{Con:98} (1998) has similar resolution and greater
sensitivity than the MGPS1 survey, but unfortunately the overlap region is
only about $20^\circ$ of longitude at either end.  The northern hemisphere
Westerbork Galactic Plane Survey at 327 MHz (Taylor \etal
\markcite{Tay:96}1996) has similar resolution and sensitivity to MGPS1 and
many of the outcomes from that survey will have a counterpart in the
present work.

This paper presents the strategies used to produce the survey in Section 2,
an outline of the reduction procedures including an assessment of the
quality and limitations of the data in Section 3, and the results from the
survey in Section 4.  MOST studies of the Galactic Centre (Gray
\markcite{Gra:94} 1994a,b,c,d) and the Vela supernova remnant (Bock, Turtle
\& Green \markcite{Boc:98} 1998) should be regarded as extensions of this
survey.

\section{Observations}


The MOST is an Earth-rotation synthesis radio telescope with the technical
parameters listed in Table 1. For the survey reported here, all
observations were made in a time-shared mode for which the field of view is
$70' \times 70' \cosec |\delta|.$ As explained by Mills
\markcite{Mil:81}(1981) and Robertson \markcite{Rob:91}(1991), the MOST is
distinguished from other Earth-rotation synthesis telescopes by two
principal differences.

Firstly, the MOST is an east-west cylindrical paraboloid reflector which
illuminates a line feed comprising nearly nine thousand ring elements.  It
tracks a point in the sky by mechanically steering the {\it tilt} of the
reflector about its long axis, and electronically and mechanically steering
the {\it meridian distance} of the cone beams formed by the feed line.
Figure 1 illustrates the tilt and meridian distance ($MD$) co-ordinates which
form a natural system for the MOST.


The combination of (slow) mechanical and (fast) electronic steering of the
cone beams, taken together with the regular spacing of the sub-structures
which comprise the feed line, leads to grating responses lying
$1.15^\circ$~sec~$(MD)$ away from the central response.  The amplitude of
these is small near the centre of the field but, if uncorrected, rises to
about 20\% of the central response at the edge of the field.  Deconvolution
of these responses has proven to be difficult, but modifications (Amy \&
Large \markcite{Amy:89} 1989) undertaken during the course of the survey
significantly reduced their amplitude. In addition to inducing grating
responses, the long, cylindrical form of the MOST also forms an
instantaneous primary beam which is not azimuthally symmetric.  This
implies that different parts of a synthesised image are subject to
different attenuation, resulting from the integrated effect of the
time-varying instantaneous primary beam. We call this attenuation the
`synthesised primary beam.' No difficulties are encountered in dealing with
its effect on the images.
 
The second distinguishing characteristic of the MOST is that it forms fan
beam responses in real time and records these directly, thus bypassing the
need to record complex visibilities. This process greatly reduces the
required computing power, but entails vector averaging over all redundant
baselines. Such averaging removes the possibility of antenna-based self
calibration. The consequences of these differences are discussed below.

With a field of view of $70' \times 70' \cosec |\delta|$, the MOST is well
suited to imaging large areas of the sky. Nevertheless, a complete survey
of the southern Galactic plane has necessitated a number of compromises
whose impact on the survey are described below.

\subsection{Observing Strategies}

Complete synthesis of MOST images requires full coverage of hour angles,
$h$, satisfying $-6^{h} \leq h \leq +6^{h}$. This is done by tracking the
field centre for 12 h, sampling the fan-beam responses every 24 s. Owing to
(intended) mechanical limitations in the MOST, complete synthesis is
possible only for declinations south of approximately $-30^\circ$. The
survey has thus been restricted to $245^\circ \leq l \leq 355^\circ$. Gray
\markcite{Gra:94} (1994a,b,c,d) has surveyed a $10^\circ \times 5^\circ$
($l \times b$) field at the Galactic centre which should be regarded as the
northern extension of the survey presented here.

Being an east-west array, the MOST forms an elliptical synthesised
interferometer beam that is elongated by a factor of cosec $|\delta|$ in
declination relative to right ascension. The field of view within the
synthesised primary beam has a similar aspect ratio.  When this fact is
combined with the arc-like projection of the Galactic plane onto the
southern sky, it implies that a rather complex grid of standard field
centres is needed to cover the survey area efficiently.  In addition, to
minimize the challenges of dealing with the grating responses described
above it was desirable to allow for considerable overlap of fields.  A
standard survey grid was adopted in which adjacent fields overlap to within
$\sim10'$ of the field centres.  Many additional observations were made
with non-standard field centres, to further reduce the impact of grating
responses and other artifacts produced in the observations positioned on
the standard grid. The surveyed area and the coverage is illustrated in
Figure 2.
 

Preferentially, MOST synthesis observations are made at night with the
field centre transiting near midnight. This minimises the effects of solar
radiation at radio frequencies and reduces exposure to artificial
interference.  Consequently, most observations were made each year between
February and June.  The complete log of the 455 observations used in the
survey is listed in Table 2, which gives the Galactic name of each field
observed, the RA and Dec. at epoch J2000 of the pointing centre, the date
of observation, and a flag (N) if the field was not part of the standard
grid. This Table will allow users of the survey to identify component
fields containing selected sources, and also facilitate searches for
variability of sources detected at different times in this survey, or in
any future work.  In cases when pointing centres were repeated, multiple
dates are given only for data actually used in the final survey.

\subsection{Primary Calibration}

The primary calibration of images in the survey rests on observations of a
sample of strong, compact sources made immediately before and after each 12
hour synthesis. These short fan-beam observations are known as `scans.'
The 843 MHz flux densities and positions of the calibration sources have
been compiled by Hunstead \markcite{Hun:91} (1991) and updated (cataloguing
known source variability) by Campbell-Wilson \& Hunstead \markcite{Cam:94}
(1994). The absolute flux density scale is believed to be accurate to $\pm
5$\%, and the positions are tied to the southern VLBI frame to an
uncertainty of about $\pm 0.5$ arcsec. The output of a scan observation of
a calibration source is an estimate of the gain of the telescope (counts
Jy$^{-1}$) and the pointing offset in meridian distance (arcsec). To reduce
the effects of confusion, intrinsic variability and scintillation on the
calibration parameters, up to 8 sources are observed before and another 8
after each synthesis observation. The gain and pointing offsets are
determined for each of these sources and a median fit to the entire set is
used to calibrate the corresponding image.

The large amount of overlap in the component fields of the survey allows
refinement of the calibration factors after the entire data set has been
analyzed. To do this, the flux densities and positions of strong,
unresolved sources detected in each image were assembled into a data base,
which collated the multiple measurements that exist for many of the
sources. Median filtering of these repeated measurements provided an
additional calibration correction for each image, which could be applied
either by re-imaging or during the assembly of the mosaics (Section 3.2).

The variation in the mean of the daily scan calibration factors is about
10\% for the gain, and 2.0$''$ for the meridian distance offset (which
leads to a positional uncertainty of about 2.5$''$ in the final image).
The use of overlapping fields reduced these to about 5\% for the final flux
density scale, and to about $1'' \times 1'' \cosec |\delta|$ in position,
for sources stronger than 20 mJy.  As we note below, the complexity and
high brightness of much Galactic emission ensures that for a significant
fraction of the survey, the image quality is not determined by calibration
uncertainties.

\section{Data Analysis}

The formation of the final data product of the survey has involved many
time-consuming and often complex computations, mainly to deal with
artefacts. While the details of these processes are not necessarily of
general interest, it is important that users of the survey appreciate all
of the significant consequences for subsequent interpretation of the
images.  Hence, we provide here a brief but fairly complete account of the
reduction procedures.

\subsection{Editing \& Auxiliary Calibration}

MOST data are corrupted on occasion by artificial radio sources
(principally mobile telephones) and by electromagnetic radiation associated
with lightning and the Sun. Typically, only 0 -- 10 samples (each of 24
solar seconds duration) are affected by artificial interference or
lightning in any 12 (sidereal) hour data set of 1796 samples, but in view
of the large amplitude of some of the interference it must be removed prior
to image formation. The corrupted samples are identified during
automated or visual inspection of the raw data and then deleted.
Solar interference has a different character, being manifest as
low-level, large scale bands aligned in directions corresponding to the
hour angles at which the Sun lies in one of the MOST's sidelobes. Despite
the complexity of the telescope response to the Sun, solar interference can
be satisfactorily removed by carefully adjusting the baselevel of the image
using Fourier techniques.

Most sample deletions can be compensated for by ensuring that the shape and
amplitude of the corresponding deconvolution beam correctly reflects their
existence. Nevertheless, in images where there is a very strong source
located just outside the field in question, it is not possible to CLEAN the
sidelobes of this source completely. A radial artefact at the hour angle of
any deleted sample will be produced. There are only a few such artefacts
remaining in the survey, most of them having been removed interactively by
using the relevant overlapped images which are free from interference at the
hour angles in question.

In addition to the primary calibration of the daily flux density scale
described above, auxiliary calibrations must be made to correct for (1)
environmental variations, (2) attenuation due to the primary beam, and (3)
variation of the telescope gain as a field is tracked in meridian distance.

The main environmental correction arises from the thermal sensitivity of
several components in the receiver chain exposed to the atmosphere.  The
temperature sensitivity of the throughput of the entire system is
determined from time to time by tests which show that the calibration
factor is well-described by a stable function of temperature. To evaluate
this function, the ambient temperature is measured and recorded with each
data sample. The uncertainty in the flux density scale due to this
calibration is approximately $\pm 1$\%. The gain of the telescope is
sometimes reduced anomalously following periods of rain, due to attenuation
in the local oscillator trough line.  The few observations badly
compromised by wet weather have been rejected and the field subsequently
repeated.

Because the MOST is a cylindrical paraboloid, the projection of the primary
beam in the image plane varies with the hour angle of the field centre.
This variation is readily modelled on the basis of telescope calibrations
performed prior to the survey, and can be easily removed in the imaging
process. Even at the declination extremes of the most northerly fields in
the survey, where the effects of the primary beam attenuation are largest,
the correction is no more than a factor of 1.2, with an uncertainty no
larger than 10\% of this value. The correction is much smaller over most of
the surveyed area.

The variation of telescope gain with meridian distance (called the
`meridian distance gain curve') consist of two parts. The first is a
monotonic decline towards large meridian distances due to the
foreshortening of the collecting area of the horizontal paraboloid.  The
second is a modulation of order $\pm 20$ \% on an angular scale of about
$10^\circ$, due to constructive and destructive interference between rays
that are directly reflected, and those scattered from the feed line and its
supports. The meridian distance gain curve of the telescope is invariant
and calibrated with an uncertainty of about 2\%. A correction for the
modulation is readily made in the imaging process.

\subsection{Image Formation}


A flow chart of the procedure used to produce the final images is shown in
Figure 3. It involves two major cycles of imaging, as well as several steps
designed to ensure a uniform calibration and to remove as many artefacts as
possible.

In both cycles, MOST fan beam responses are converted to an image using the
back projection algorithm (Perley \markcite{Per:79}1979,
Crawford \markcite{Cra:84}1984).  The algorithm
correctly forms the image in the selected co-ordinate system, applying
appropriate corrections for precession and small geodetic misalignments in
the baseline of the MOST. The images are formed initially in equatorial
coordinates and the NCP projection (see Greisen \markcite{Gre:83}1983) to
facilitate deconvolution.  As a result of the extensive ($u,v$) coverage of
a MOST observation, the synthesised point spread function is compact and
relatively simple. Its first sidelobe has an amplitude of $-8$\%, while the
distant sidelobes all have amplitudes $<1$\%.

The images are CLEANed using an image-plane algorithm similar to that
described by Hogbom \markcite{Hog:74} (1974).  A source fitting algorithm
(Crawford \markcite{Cra:89}1989) is then applied to provide a list of all
of the compact sources. The list is sorted and amalgamated for the entire
survey, and then used for (1) the modelling and removal of grating
responses, (2) the iterative refinement of calibration parameters as
described above, and (3) the production of a source catalogue to be
presented in Paper II (in preparation).

The second and final synthesis of individual fields includes adjustment of
the calibration factors, amelioration of grating rings (as an extension of
the CLEAN process), the excision by Fourier filtering of any low-level
solar interference, and the adaptive deconvolution of an image provided
there is a strong, compact source ($S_{843} \ge 200$ mJy) near the centre
of the field (Cram \& Ye \markcite{Cra:95} 1995).  Typically, the
calibration corrections at this final stage may change flux densities by
$\sim 2\%$ and positions by $< 1 ''$.

The last step in forming survey images is to combine individual fields into
a $3^{\circ} \times 3^{\circ}$ mosaic in Galactic co-ordinates.  Each
$3^{\circ} \times 3^{\circ}$ mosaic contains at least 16 component images.
Before combination, every image is inspected visually, and areas containing
evident artefacts are flagged for possible exclusion. After all component
images have been flagged, the mosaic is assembled. Sometimes, a flagged
artefact will appear at the same position in all of the existing
observations. The flagging process will then create a gap and the images
are re-examined to select the least affected for inclusion.

\subsection{Image Quality \& Residual Artefacts}

The images themselves (Figure 4) suggest that the steps summarised above
have been very successful in providing high quality data in those parts of
the Galactic plane devoid of bright, extended sources. In regions with
bright, complex, spatially extended emission the images have a lower
dynamic range. They are still of considerable value by virtue of their good
sensitivity to low surface brightness and their relatively high angular
resolution.


The residual artefacts in MOST images originate in several ways, and in
general they can be readily distinguished from the sky emission by virtue
of their position and structure. There are three {\it major} artefacts to
consider, arising from (a) propagation effects, (b) grating responses, and
(c) missing short spacings.  The region of the sky near $l=284.0^{\circ}$
[Figure 4(i)] illustrates one of the most difficult areas to image, and all
three types of residual artefacts can be discerned.  In addition, the
projection of images to form the mosaic in $(l,b)$ coordinates distorts the
original synthesised beam shape, and the published images therefore have a
position-dependent beam.  We now consider each of these artefacts in turn.

\subsubsection{Propagation effects}

Propagation effects arise through time-dependent refractive gradients in
the ionosphere and troposphere. These phenomena lead to bright and dark
patterns which radiate from high brightness sources. The patterns are the
direct consequence of the time-dependent distortion of the wavefront on its
passage through the atmosphere, which this leads to small ($\sim 1-2''$
amplitude), time-dependent ($\sim 10$ min time scales) apparent
displacements in the positions of sources.  Adaptive deconvolution (Cram \&
Ye \markcite{Cra:95} 1995) has partially removed the effect but has been
unable to improve the dynamic range beyond a ratio of $250:1$.

\subsubsection{Grating responses}

The grating responses of the fan beams of the MOST are manifest in images
as elliptical rings or arcs around bright sources, with minor and major
radii $1.15^{\circ}n$ and $1.15^{\circ}n$ cosec $|\delta|$, respectively.
Here, $n$ is the order of the response.  Most of the residual artefacts
correspond to $n=+1$, although for sources stronger than about 2.5 Jy a
second and even a third ellipse may be detected. As explained by Amy \&
Large \markcite{Amy:89}(1989), the amplitude of the grating response of the
fan beams changes with the meridian distance of the field, and hence the
amplitude of the synthesised ring varies across the field of view.  In
particular, the response is very weak near the field centre (i.e., from
sources lying $1.15^{\circ}$ from a field centre).  Considerable efforts
(involving real-time signal processing) were expended to reduce these
responses as MGPS1 progressed.  While these were successful, the cost is a
residual response which is complex and therefore difficult to remove fully.
The problem is exacerbated when the grating responses are produced by
sources that are strong and complex.

\subsubsection{Missing short spacings}

The MOST does not measure the auto-correlation of its elements, nor does it
measure correlations on baselines shorter than the inter-arm gap of $42.9
\lambda$. Hence, structures with angular sizes larger than $\sim 20' - 30'$
are not detected and in unCLEANed images all sources are surrounded by a
`bowl' of apparently negative flux density. The CLEAN algorithm can go some
way to interpolating the unmeasured short spacings and thus to restoring
the negative bowl (e.g.  Schwarz \markcite{Sch:84}1984). However, small
residual uncertainties in our knowledge of the actual MOST beam preclude
the complete removal of this artefact in the presence of bright, extended
sources that contain more than $\approx 100$ Jy of flux density. An
artefact related to the negative bowl arises when images containing 
such sources overlap at the mosaic. In this case,
differences in the telescope beam and the depth of CLEANing in the two
images may lead to systematic offsets between the local image base levels,
manifest as a small, sharp step in the mosaiced image.

\subsubsection{Position dependent synthesised beam}

When the individual images comprising the survey were restored following
the CLEAN process, the restoring beam was independent of position in the
NCP projection. However, different images comprising a given mosaic have
been restored with different beams, since the beam dimensions are set by
the declination of the field centre (i.e., the phase centre).
Consequently, the beam shape at any point in a MGPS1 mosaic is a linear
combination of the set of beams for all the images contributing at that
point (Sault, Killeen \& Brouw \markcite{Sau:96} 1996). This beam shape has
been further distorted by the process of regridding to $(l,b)$ coordinates.
As a result, it is entirely inappropriate for users of the survey to
attempt to deconvolve sources in the mosaics. Source fitting, especially of
compact sources, should be done on the original component images.

\section{Results and prospects}


The working product of the survey is a set of 37 mosaics, each covering an
area of $3^{\circ} \times 3^{\circ}$ in Galactic co-ordinates. Each mosaic
is composed of at least 16 calibrated individual fields, with simple
averaging in overlapped regions.  In this paper the images are further
amalgamated into $9^{\circ} \times 3^{\circ}$ mosaics, shown in Figures
4(a) -- 4(m), as greyscales within the range (white) $-30$ to (black) $+40$
mJy beam$^{-1}$. The peak amplitude of the stronger sources seen in the
survey is often above 1 Jy, and they are thus burnt out in the images.
However, the selected grayscale range exemplifies the capacity of the MOST
to reveal faint, large-scale structure.

The mosaics show a panorama of the Galactic Plane characterised by a host
of unresolved and small-diameter sources, together with a highly
non-uniform population of extended and complex sources.  Most of the
small-diameter objects are background galaxies and their distribution
appears to be uniform ( Whiteoak \markcite{\Whi:92} 1992).  There are also
many extended objects which exhibit the typical morphology of radio
galaxies (e.g., double-lobe, head-tail and wide-angle tail sources)
distributed without any obvious clustering throughout the survey region,
and probably all of these will also be background sources.
Multi-wavelength studies of these extragalactic sources will help improve
understanding of the sky in the Zone of Avoidance.

By contrast, there is a marked tendency for the host of large Galactic
sources seen in the images to be apparently clustered into complexes.  The
most prominent of these complexes tend to occur at the longitudes
associated with tangent directions to Galactic spiral arms (Mills
\markcite{Mil:59} 1959; Green \markcite{Gre:74} 1974) and may well
represent superpositions of several sources along the line of sight.  The
effect is especially obvious along the tangent direction at longitudes
$312^{\circ}$ [Figure 4(f)] and $332^{\circ}$ [Figure 4(d)].  However,
there are other complexes lying in directions not known to be associated
with tangent point directions. Some of these have morphologies suggestive
of large scale physical association within the complex. For example, there
are large plumes and groupings which strongly suggest interactions at
longitudes $317^{\circ}$ [Figure 4(e)] and $327^{\circ}$ [Figure 4(d)].
The images show many filaments extending over several degrees. It is not
known whether these structures have a separate physical identity or are
perturbations in the emission from larger, more homogeneous phases of the
interstellar medium (ISM).

Multi-wavelength comparisons can be made between MGPS1 and the IRAS 60
$\mu$m survey. Thermal sources dominate the IRAS images and the ratio of
the flux densities at 60 $\mu$m and 843 MHz proves to be a powerful
discriminant between thermal and non-thermal processes (Whiteoak \& Green
\markcite{Whi:96} 1996; Taylor \etal \markcite{Tay:96}1996).  Further
insights will be gained when these data are related to the wide-field
H$\alpha$ imaging surveys currently in progress at the Siding Spring
Observatory (Buxton, Bessel \& Watson \markcite{Bux:98}1998; Parker
\& Phillips \markcite{Mas:98}1998).  Detailed correlations can also
be explored with the overlapping portion of the 1.4 GHz NRAO-VLA Sky Survey
(Condon \etal \markcite{Con:98} 1998), and with the 5 GHz Parkes-MIT-NRAO
survey (Griffith \& Wright \markcite{Gri:93} 1993).

The MGPS1 can potentially reveal {\it variable} radio sources, from
inspection of the regions of overlap between the $70'$ images. Any such
variability will be detected in the process of developing a source
catalogue, which is currently in preparation. Furthermore, a larger second
epoch survey, MGPS2, is currently in progress to explore variability on
longer time scales and to extend the surveyed region to higher galactic
latitudes (Large \etal \markcite{Lar:94}1994; Green \markcite{Gre:97}1997,
\markcite{Gre:98}1998).  It is hoped the MGPS1 will form the basis for many
individual source investigations as well as studies of large-scale
structures in the ISM, including identification of relic SNRs and
wind-blown bubbles of ionised gas.  To assist in this work, $3 \times
3^\circ$ versions of the images displayed in this paper (together with
subimages of selected objects, such as the supernova remnants described by
Whiteoak \& Green 1996) are available in FITS format from the website
http://www.astrop.physics.usyd.edu.au/MGPS. The original images, necessary
for quantitative source fitting, may also be downloaded from this site.

\acknowledgements The Molonglo Observatory Site Manager,
D.~Campbell-Wilson, and the technical staff, J.~Webb, M.~White and
B.~Smithers have made a considerable contribution to the success of this
project and we warmly acknowledge their dedication and skill.
B.~Piestrzynski analysed and organised the data archive and managed the
data processing with outstanding care and commitment.  Software used in the
data reduction has been developed with assistance from several astronomers,
including J.~Reynolds, S.~Amy, A.~Gray and V.~McIntyre, and we gratefully
acknowledge their contribution. The MOST is operated with the support of
the Australian Research Council and the Science Foundation for Physics
within the University of Sydney.

\clearpage

\figcaption[FIG1]{Definition of the tilt and meridian distance ($MD$)
  co-ordinates of a point $P$. This is the natural system for the MOST. The
  directions $N$ and $W$ point to North and West, respectively, and $Z$ is
  towards the Zenith.}

\figcaption[FIG2]{Plot showing the location of all the component fields
  observed as part of MGPS1. Pointing centres associated with the Galactic
  Centre (Gray 1994a,b,c,d) and Vela (Bock \etal~1998) complementary
  surveys are also shown.}

\figcaption[FIG3]{Flow chart showing the process used to produce the final
  product of the survey, namely $3^\circ\times 3^\circ$ mosaics in Galactic
  co-ordinates.}

\figcaption[FIG4]{Large mosaics, $9^\circ\times 3^\circ (l \times b)$, of
  the Galactic plane covered by MGPS1. The greyscale has been compressed
  to the range -30 to +40 mJy beam$^{-1}$ to show more clearly the faint,
  extended structure. The full dynamic range is available in the electronic
  version of the data.}

\clearpage

\begin{table}
\caption{Technical specifications of the MOST}
\begin{tabular}{ll}
  \hline Centre frequency & 843 MHz ($\lambda$ = 36 cm) \\ 
  Bandwidth & 3 MHz \\ 
  Polarization & RHC (IEEE) \\ 
  Declination range & -90$^\circ$ to $\approx -30^\circ$ \\ 
  Synthesised beam & $43'' \times 43'' \cosec |\delta|$ \\ 
  Field size (MGPS1) & $70' \times 70' \cosec |\delta|$ \\ 
  Noise after 12 h & 1 - 2 mJy beam $^{-1}$ (1~$\sigma$) \\ 
  Surface brightness sensitivity & $\approx$ 1 K ($1~\sigma$) \\ 
  Dynamic range (best) & 250:1 \\ \hline
\end{tabular}
\end{table}



\clearpage
\begin{center}
\begin{minipage}[t]{10cm}
  \psfig{figure=f1.epsi,width=10cm,angle=0,clip=}
\end{minipage}
\end{center}
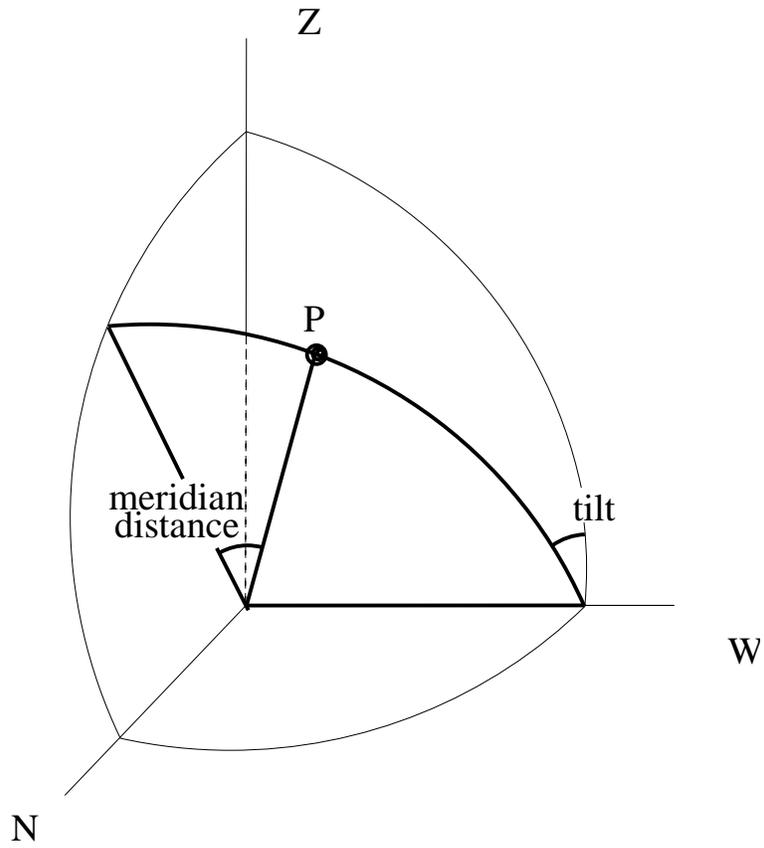
Figure 1. Definition of the tilt and meridian distance co-ordinates of a
point $P$. This is the natural system for the MOST.  $N$ and $W$ point to
North and West, respectively, and $Z$ is the Zenith.

\clearpage
\begin{center}
\begin{minipage}[t]{15cm}
  \psfig{figure=f2.epsi,width=15cm,bbllx=-300pt,bblly=0pt, bburx=600pt,
    bbury=800pt,angle=-90,clip=}
\end{minipage}
\end{center}
Figure 2. Plot showing the location of all the component fields observed as
part of MGPS1.  Pointing centres associated with the Galactic Centre (Gray
1994a,b,c,d) and Vela (Bock \etal 1998) complementary surveys are also
shown.

\clearpage
\begin{center}
\begin{minipage}[t]{14cm}
  \psfig{figure=f3.epsi,width=14cm,clip=} 
\end{minipage}
\end{center}
Figure 3. Plot showing the location of all the component fields observed as
part of MGPS1.  Pointing centres associated with the Galactic Centre (Gray
1994a,b,c,d) and Vela (Bock \etal 1998) complementary surveys are also
shown.

\end{document}


\setcounter{table}{1}
\newcommand{\phz}{\phm{0000}}

\begin{deluxetable}{lcclllccll}

\footnotesize
\tablecaption{Log of Observations of MGPS--1}
\tablewidth{0pt}
\tableheadfrac{0.06}
\tablehead
{
\colhead{Field Centre} & 
\colhead{$\alpha$(J2000)} &  
\colhead{$\delta$(J2000)} & 
\colhead{Date} & 
\colhead{} &
\colhead{Field Centre} & 
\colhead{$\alpha$(J2000)} &  
\colhead{$\delta$(J2000)} & 
\colhead{Date} & 
\colhead{}
}

\startdata
G245.25$-$0.9  & 07 52 17.2   &  $-$28 52 32   &  1990 Jan 30  &    \phz &      G303.75$+$0.0   &  12 58 36.7 & $-$62 51 37 & 1989 Apr 14  &    \nl
G245.25$+$0.9  &  07 59 17.2   &  $-$27 56 33   &  1990 Jan 14  &    \phz &     G303.8$-$0.7   &  12 59 07.0 & $-$63 36 11 & 1991 Apr 19  &  N \nl
             &               &               &  1991 Sep 28  &    \phz &        G304.12$-$0.8  &  13 02 08.9 & $-$63 41 08 & 1990 Mar 14  &  N \nl
G245.75$+$0.0  &  07 56 59.1   &  $-$28 50 20   &  1987 Feb 02  &    \phz &     G304.125$-$0.9 &  13 02 13.7 & $-$63 44 48 & 1989 Apr 16  &    \nl
G246.25$-$0.9  &  07 54 38.6   &  $-$29 43 59   &  1990 Jan 15  &    \phz &     G304.125$+$0.9  &  13 01 35.2 & $-$61 56 54 & 1989 Apr 30  &    \nl
G246.25$+$0.9  &  08 01 40.8   &  $-$28 47 32   &  1990 Jan 16  &    \phz &     G304.5$+$0.0    &  13 05 10.8 & $-$62 49 48 & 1991 Apr 14  &    \nl
             &               &               &  1992 Jan 31  &    \phz &        G304.875$-$0.9 &  13 08 59.7 & $-$63 42 20 & 1989 Apr 19  &    \nl
G246.75$+$0.0  &  07 59 22.9   &  $-$29 41 29   &  1989 Jan 17  &    \phz &     G304.875$+$0.9  &  13 07 57.2 & $-$61 54 34 & 1992 May 01  &    \nl
G247.25$-$0.9 &  07 57 02.5   &  $-$30 35 17   &  1991 Sep 29  &    \phz &      G305.25$+$0.0   &  13 11 43.8 & $-$62 46 49 & 1983 Apr 18  &    \nl
G247.25$+$0.9  &  08 04 06.9   &  $-$29 38 22   &  1991 Sep 08  &    \phz &     G305.625$-$0.9 &  13 15 44.2 & $-$63 38 41 & 1986 Apr 09  &    \nl 
G247.75$+$0.0  &  08 01 49.2   &  $-$30 32 29   &  1987 Feb 04  &    \phz &     G305.625$+$0.9  &  13 14 18.0 & $-$61 51 08 & 1993 Mar 22  &    \nl
G248.25$-$0.9 &  07 59 28.9   &  $-$31 26 26   &  1991 Sep 22  &    \phz &      G305.9$+$1.0    &  13 16 58.7 & $-$61 45 49 & 1991 Apr 03  &  N \nl
G248.25$+$0.9  &  08 06 35.5   &  $-$30 29 01   &  1991 Oct 04  &    \phz &     G306.0$+$0.0    &  13 18 15.3 & $-$62 42 43 & 1993 Apr 03  &    \nl
G248.75$+$0.0  &  08 04 18.0   &  $-$31 23 17   &  1991 Nov 01  &    \phz &     G306.375$-$0.9 &  13 22 26.7 & $-$63 33 51 & 1989 Apr 23  &    \nl
G249.25$-$0.9 &  08 01 57.8   &  $-$32 17 25   &  1991 Sep 21  &    \phz &      G306.375$+$0.9  &  13 20 37.1 & $-$61 46 35 & 1989 May 07  &    \nl
G249.25$+$0.9  &  08 09 06.8   &  $-$31 19 28   &  1991 Oct 12  &    \phz &     G306.4$+$0.1    &  13 21 17.4 & $-$62 35 43 & 1992 Jun 26  &  N \nl
G249.75$+$0.0  &  08 06 49.6   &  $-$32 13 55   &  1991 Nov 02  &    \phz &     G306.75$+$0.0   &  13 24 44.7 & $-$62 37 29 & 1989 May 03  &    \nl
G250.25$-$0.9 &  08 04 29.5   &  $-$33 08 13   &  1991 Sep 27  &    \phz &      G307.125$-$0.9 &  13 29 06.7 & $-$63 27 53 & 1990 Mar 18  &    \nl
G250.25$+$0.9  &  08 11 40.9   &  $-$32 09 44   &  1991 Oct 13  &    \phz &     G307.125$+$0.9  &  13 26 54.1 & $-$61 40 58 & 1984 Apr 02  &    \nl
G250.75$+$0.0  &  08 09 23.8   &  $-$33 04 22   &  1991 Nov 23  &    \phz &     G307.5$+$0.0    &  13 31 11.5 & $-$62 31 07 & 1983 Apr 20  &    \nl
G251.25$-$0.9 &  08 07 04.1   &  $-$33 58 50   &  1991 Oct 06  &    \phz &      G307.875$-$0.9 &  13 35 43.7 & $-$63 20 45 & 1990 Mar 20  &    \nl
G251.25$+$0.9  &  08 14 17.9   &  $-$32 59 47   &  1991 Oct 18  &    \phz &     G307.875$+$0.9  &  13 33 08.6 & $-$61 34 15 & 1990 Mar 19  &    \nl
             &               &               &  1991 Nov 15  &    \phz &        G308.25$+$0.0   &  13 37 35.4 & $-$62 23 39 & 1986 Apr 07  &    \nl
G251.75$+$0.0  &  08 12 01.3   &  $-$33 54 36   &  1991 Nov 24  &    \phz &     G308.625$-$0.9 &  13 42 17.2 & $-$63 12 30 & 1988 Apr 14  &    \nl
G252.25$-$0.9 &  08 09 41.8   &  $-$34 49 15   &  1991 Jan 07  &    \phz &      G308.625$+$0.9  &  13 39 20.2 & $-$61 26 28 & 1983 Apr 21  &    \nl
G252.25$+$0.9  &  08 16 57.9   &  $-$33 49 37   &  1991 Jan 09  &    \phz &     G309.0$+$0.0    &  13 43 55.9 & $-$62 15 06 & 1983 Apr 26  &    \nl
G252.75$+$0.0  &  08 14 41.8   &  $-$34 44 37   &  1991 Nov 16  &    \phz &     G309.2$-$0.4   &  13 46 30.3 & $-$62 35 00 & 1991 Apr 11  &  N \nl
G253.25$-$0.9 &  08 12 22.6   &  $-$35 39 27   &  1990 Jan 17  &    \phz &      G309.375$-$0.9 &  13 48 46.7 & $-$63 03 08 & 1985 Apr 11  &    \nl
G253.25$+$0.9  &  08 19 41.2   &  $-$34 39 14   &  1991 Jan 15  &    \phz &     G309.375$+$0.9  &  13 45 28.5 & $-$61 17 38 & 1986 May 01  &    \nl
G253.75$+$0.0  &  08 17 25.5   &  $-$35 34 25   &  1992 Mar 29  &    \phz &     G309.7$+$1.7    &  13 46 45.5 & $-$60 24 59 & 1994 Mar 28  &  N \nl
G254.25$-$0.9 &  08 15 06.7   &  $-$36 29 27   &  1990 Jan 18  &    \phz &      G309.75$+$0.0   &  13 50 12.5 & $-$62 05 29 & 1985 Feb 02  &    \nl
G254.25$+$0.9  &  08 22 27.8   &  $-$35 28 36   &  1991 Jan 19  &    \phz &     G310.0$+$0.5    &  13 51 30.4 & $-$61 34 50 & 1991 Apr 15  &  N \nl
G254.75$+$0.0  &  08 20 12.7   &  $-$36 23 59   &  1984 Feb 02  &    \phz &     G310.125$-$0.9 &  13 55 11.8 & $-$62 52 40 & 1988 Apr 24  &    \nl
G255.25$-$0.9 &  08 17 54.3   &  $-$37 19 13   &  1991 Oct 20  &    \phz &      G310.125$+$0.9  &  13 51 33.1 & $-$61 07 46 & 1988 apr 25  &    \nl
G255.25$+$0.9  &  08 25 17.9   &  $-$36 17 42   &  1991 Aug 11  &    \phz &     G310.5$+$0.0    &  13 56 25.0 & $-$61 54 48 & 1985 Apr 15  &    \nl
G255.75$+$0.0  &  08 23 03.4   &  $-$37 13 18   &  1986 Feb 10  &    \phz &     G310.8$+$0.0    &  13 59 05.5 & $-$62 10 34 & 1991 Apr 16  &  N \nl
G256.25$-$0.9 &  08 20 45.6   &  $-$38 08 44   &  1991 Jan 17  &    \phz &      G310.875$-$0.9 &  14 01 32.2 & $-$62 41 09 & 1992 Jun 12  &    \nl
G256.25$+$0.9  &  08 28 11.6   &  $-$37 06 33   &  1991 Aug 25  &    \phz &     G310.875$+$0.9  &  13 57 33.7 & $-$60 56 52 & 1986 May 07  &    \nl
G256.75$+$0.0  &  08 25 57.9   &  $-$38 02 21   &  1986 Feb 11  &    \phz &     G311.1$+$0.2    &  14 00 31.6 & $-$61 34 41 & 1990 Apr 05  &  N \nl
G257.25$-$0.9 &  08 23 40.7   &  $-$38 58 00   &  1991 Aug 10  &    \phz &      G311.25$+$0.0   &  14 02 33.0 & $-$61 43 05 & 1993 Apr 09  &    \nl
G257.25$+$0.9  &  08 31 09.2   &  $-$37 55 07   &  1991 Aug 24  &    \phz &     G311.625$-$0.9 &  14 07 47.4 & $-$62 28 34 & 1985 Apr 17  &    \nl
G257.75$+$0.0  &  08 28 56.3   &  $-$38 51 08   &  1986 Feb 12  &    \phz &     G311.625$+$0.9  &  14 03 30.1 & $-$60 44 59 & 1985 Apr 18  &    \nl
G258.25$-$0.9 &  08 26 39.9   &  $-$39 47 00   &  1990 Jan 19  &    \phz &      G312.0$-$3.0   &  14 20 52.1 & $-$64 13 44 & 1995 Sep 02  &  N \nl
G258.25$+$0.9  &  08 34 10.8   &  $-$38 43 24   &  1992 Feb 03  &    \phz &     G312.0$+$0.0    &  14 08 36.1 & $-$61 30 21 & 1984 Apr 03  &    \nl
G258.75$+$0.0  &  08 31 58.8   &  $-$39 39 38   &  1986 Feb 13  &    \phz &                   &             &            & 1990 Apr 15  &    \nl
G259.25$-$0.9 &  08 29 43.2   &  $-$40 35 42   &  1989 Jan 19  &    \phz &      G312.375$-$0.9 &  14 13 57.3 & $-$62 14 57 & 1988 Apr 26  &    \nl
G259.25$+$0.9  &  08 37 16.6   &  $-$39 31 22   &  1989 Jan 18  &    \phz &     G312.375$+$0.9  &  14 09 21.9 & $-$60 32 06 & 1991 Apr 05  &    \nl
G259.75$+$0.0  &  08 35 05.7   &  $-$40 27 49   &  1986 Feb 18  &    \phz &     G312.5$-$0.2   &  14 13 04.3 & $-$61 34 13 & 1990 Apr 09  &  N \nl
G260.25$-$0.9 &  08 32 51.0   &  $-$41 24 07   &  1992 Mar 01  &    \phz &      G312.75$+$0.0   &  14 14 34.1 & $-$61 16 37 & 1986 Apr 30  &    \nl
G260.25$+$0.9  &  08 40 26.8   &  $-$40 19 00   &  1994 Jan 26  &    \phz &     G313.125$-$0.9 &  14 20 01.4 & $-$62 00 21 & 1990 Apr 11  &    \nl
G260.75$+$0.0  &  08 38 17.0   &  $-$41 15 41   &  1986 Feb 20  &    \phz &     G313.125$+$0.9  &  14 15 08.8 & $-$60 18 16 & 1990 Apr 10  &    \nl
G261.25$-$0.9 &  08 36 03.5   &  $-$42 12 13   &  1988 Feb 19  &    \phz &                    &             &            & 1992 Mar 25  &    \nl
G261.25$+$0.9  &  08 43 41.5   &  $-$41 06 18   &  1984 Feb 03  &    \phz &     G313.5$+$0.0    &  14 20 26.8 & $-$61 01 56 & 1990 Apr 29  &    \nl
G261.75$+$0.0  &  08 41 33.1   &  $-$42 03 13   &  1989 Jan 20  &    \phz &     G313.875$-$0.9 &  14 25 59.6 & $-$61 44 46 & 1990 Apr 14  &    \nl
G262.25$-$0.9 &  08 39 20.8   &  $-$42 59 58   &  1988 Feb 21  &    \phz &      G313.875$+$0.9  &  14 20 50.7 & $-$60 03 30 & 1990 Apr 28  &    \nl
G262.25$+$0.9  &  08 47 01.1   &  $-$41 53 14   &  1988 Feb 20  &    \phz &     G314.25$+$0.0   &  14 26 13.9 & $-$60 46 18 & 1993 Apr 18  &    \nl
G262.75$-$1.8 &  08 37 03.8   &  $-$43 56 34   &  1992 Jan 29  &  N \phz &      G314.625$-$0.9 &  14 31 51.6 & $-$61 28 15 & 1990 Apr 30  &    \nl
G262.75$+$0.0  &  08 44 54.1   &  $-$42 50 23   &  1992 Feb 12  &    \phz &     G314.625$+$0.9  &  14 26 27.4 & $-$59 47 49 & 1991 Apr 17  &    \nl
G263.0$-$0.2  &  08 44 53.5   &  $-$-43 10 58  &  1990 Feb 09  &  N \phz &      G315.0$+$0.0    &  14 31 55.2 & $-$60 29 45 & 1993 Apr 13  &    \nl
G263.25$-$0.9 &  08 42 43.1   &  $-$43 47 23   &  1992 Feb 07  &    \phz &      G315.375$-$0.9 &  14 37 37.4 & $-$61 10 48 & 1990 apr 19  &    \nl
G263.25$+$0.9  &  08 50 25.6   &  $-$42 39 48   &  1988 Feb 22  &    \phz &     G315.375$+$0.9  &  14 31 58.6 & $-$59 31 15 & 1990 May 01  &    \nl
G263.75$+$0.0  &  08 48 20.2   &  $-$43 37 11   &  1992 Feb 15  &    \phz &     G315.4$-$2.4   &  14 42 56.2 & $-$62 29 45 & 1994 Apr 29  &  N \nl
G264.25$-$0.9 &  08 46 10.8   &  $-$44 34 25   &  1992 Feb 14  &    \phz &      G315.5$-$0.1   &  14 35 46.5 & $-$60 23 05 & 1992 Dec 06  &  N \nl
G264.25$+$0.9  &  08 53 55.4   &  $-$43 25 58   &  1989 Jan 27  &    \phz &     G315.75$+$0.0   &  14 37 30.7 & $-$60 12 18 & 1985 Feb 01  &    \nl
G264.75$+$0.0  &  08 51 51.8   &  $-$44 23 35   &  1992 Feb 16  &    \phz &     G316.125$-$0.9 &  14 43 16.7 & $-$60 52 27 & 1990 May 04  &    \nl
G265.25$-$0.9 &  08 49 44.1   &  $-$45 21 04   &  1989 Jan 29  &    \phz &      G316.125$+$0.9  &  14 37 24.3 & $-$59 13 48 & 1990 May 02  &    \nl
G265.25$+$0.9  &  08 57 30.6   &  $-$44 11 42   &  1984 Feb 07  &    \phz &     G316.5$+$0.0    &  14 43 00.0 & $-$59 53 59 & 1993 Apr 15  &    \nl
G265.75$+$0.0  &  08 55 29.0   &  $-$45 09 34   &  1989 Jan 29  &    \phz &     G316.875$-$0.9 &  14 48 49.4 & $-$60 33 15 & 1990 May 06  &    \nl
G266.25$-$0.9  & 08 53 23.2   &  $-$46 07 18   &  1989 Jan 31  &    \phz &                    &             &            & 1993 Apr 14  &    \nl
G266.25$+$0.9   & 09 01 11.5   &  $-$44 57 00   &  1989 Jan 30  &    \phz &     G316.875$+$0.9  &  14 42 44.3 & $-$58 55 32 & 1990 May 03  &    \nl
G266.75$+$0.0   & 08 59 12.1   &  $-$45 55 07   &  1993 Mar 14  &    \phz &     G316.9$-$0.1   &  14 45 18.5 & $-$59 49 38 & 1994 Apr 18  &  N \nl
G267.25$-$0.9  & 08 57 08.5   &  $-$46 53 05   &  1990 Jan 21  &    \phz &      G317.0$+$0.1    &  14 46 18.2 & $-$59 32 36 & 1990 Apr 20  &  N \nl
G267.25$+$0.9   & 09 04 58.3   &  $-$45 41 49   &  1990 Jan 22  &    \phz &     G317.2$+$1.2    &  14 43 44.4 & $-$58 32 43 & 1991 Apr 08  &  N \nl
G267.75$+$0.0   & 09 03 01.3   &  $-$46 40 11   &  1990 Jan 23  &    \phz &     G317.25$+$0.0   &  14 48 23.3 & $-$59 34 50 & 1984 Apr 04  &    \nl
G268.25$-$0.9  & 09 01 00.1   &  $-$47 38 25   &  1992 Mar 02  &    \phz &                    &             &            & 1993 Apr 16  &    \nl
G268.25$+$0.9   & 09 08 51.3   &  $-$46 26 10   &  1990 Jan 24  &    \phz &     G317.625$-$0.9 &  14 54 15.5 & $-$60 13 13 & 1990 May 11  &    \nl
G268.75$+$0.0   & 09 06 57.0   &  $-$47 24 47   &  1990 Jan 25  &    \phz &     G317.625$+$0.9  &  14 47 58.6 & $-$58 36 26 & 1990 May 05  &    \nl
G269.25$-$0.9  & 09 04 58.4   &  $-$48 23 16   &  1985 Jan 31  &    \phz &      G318.0$+$0.0    &  14 53 40.3 & $-$59 14 53 & 1990 May 10  &    \nl
G269.25$+$0.9   & 09 12 50.7   &  $-$47 09 59   &  1990 Jan 27  &    \phz &                   &             &            & 1993 Apr 30  &    \nl
G269.75$+$0.0   & 09 10 59.4   &  $-$48 08 51   &  1990 Feb 28  &    \phz &     G318.3$-$0.1   &  14 55 20.5 & $-$59 14 09 & 1992 Mar 30  &  N \nl
G270.25$-$0.9  & 09 09 03.7   &  $-$49 07 35   &  1990 Feb 01  &    \phz &      G318.3$+$0.1    &  14 55 19.9 & $-$59 02 09 & 1990 Apr 21  &  N \nl
G270.25$+$0.9   & 09 16 56.8   &  $-$47 53 16   &  1991 Feb 28  &    \phz &     G318.375$-$0.9 &  14 59 35.0 & $-$59 52 22 & 1990 May 13  &    \nl
G270.75$+$0.0   & 09 15 08.7   &  $-$48 52 22   &  1991 Feb 05  &    \phz &     G318.375$+$0.9  &  14 53 07.1 & $-$58 16 32 & 1990 May 07  &    \nl
G271.125$-$0.9 &  09 12 44.3  &  $-$49 45 56   &  1991 Feb 27  &    \phz &      G318.75$+$0.0   &  14 58 51.1 & $-$58 54 08 & 1986 May 11  &    \nl
G271.125$+$0.9  &  09 20 37.9  &  $-$48 30 39   &  1991 Feb 12  &    \phz &     G319.0$+$0.5    &  14 59 34.1 & $-$58 17 56 & 1993 May 21  &  N \nl
G271.5$+$0.0    &  09 18 20.5  &  $-$49 24 38   &  1991 Jul 28  &    \phz &     G319.1$-$0.1   &  15 01 51.4 & $-$58 46 49 & 1991 May 10  &  N \nl
G271.875$-$0.9 &  09 15 58.0  &  $-$50 18 26   &  1991 Feb 07  &    \phz &      G319.125$-$0.9 &  15 04 47.7 & $-$59 30 44 & 1990 May 14  &    \nl
G271.875$+$0.9  &  09 23 51.8  &  $-$49 02 21   &  1991 Jun 22  &    \phz &     G319.125$+$0.9  &  14 58 09.7 & $-$57 55 53 & 1990 May 12  &    \nl
G272.25$+$0.0   &  09 21 36.6  &  $-$49 56 34   &  1991 Feb 15  &    \phz &     G319.5$+$0.0    &  15 03 55.7 & $-$58 32 37 & 1993 May 01  &    \nl
G272.625$-$0.9 &  09 19 16.2  &  $-$50 50 37   &  1991 Feb 11  &    \phz &      G319.875$-$0.9 &  15 09 53.8 & $-$59 08 21 & 1988 May 13  &    \nl
G272.625$+$0.9  &  09 27 09.9  &  $-$49 33 41   &  1991 Feb 20  &    \phz &     G319.875$+$0.9  &  16 21 03.9 & $-$49 57 14 & 1984 Apr 05  &    \nl
G273.0$+$0.0    &  09 24 56.9  &  $-$50 28 08   &  1991 Feb 19  &    \phz &     G320.25$+$0.0   &  15 08 53.9 & $-$58 10 22 & 1992 Mar 28  &    \nl
G273.375$-$0.9 &  09 22 38.8  &  $-$51 22 26   &  1991 Feb 17  &    \phz &      G320.4$-$1.2   &  15 14 36.6 & $-$59 05 09 & 1990 Jun 02  &  N \nl
G273.375$+$0.9  &  09 30 32.2  &  $-$50 04 39   &  1991 Feb 23  &    \phz &     G320.5$-$1.5   &  15 16 28.1 & $-$59 21 03 & 1992 Sep 26  &  N \nl
G273.75$+$0.0   &  09 28 21.8  &  $-$50 59 21   &  1991 Feb 21  &    \phz &     G320.625$-$0.9 &  15 14 53.1 & $-$58 45 14 & 1985 May 08  &    \nl
G274.125$-$0.9 &  09 26 06.1  &  $-$51 53 54   &  1984 Feb 15  &    \phz &      G320.625$+$0.9  &  15 07 57.2 & $-$57 12 22 & 1987 May 01  &    \nl
G274.125$+$0.9  &  09 33 59.0  &  $-$50 35 14   &  1990 Feb 07  &    \phz &     G320.7$-$1.8   &  15 18 59.2 & $-$59 26 54 & 1993 Mar 29  &  N \nl
G274.5$+$0.0    &  09 31 51.3  &  $-$51 30 11   &  1990 Feb 05  &    \phz &     G320.7$+$1.1    &  15 07 48.2 & $-$57 01 30 & 1992 Mar 27  &  N \nl
G274.875$-$0.9 &  09 29 38.3  &  $-$52 24 59   &  1986 Mar 03  &    \phz &      G320.8$-$1.6   &  15 19 18.7 & $-$59 14 53 & 1992 Apr 06  &  N \nl
G274.875$+$0.9  &  09 37 30.4  &  $-$51 05 26   &  1990 Feb 08  &    \phz &     G321.0$+$0.0    &  15 13 45.9 & $-$57 47 25 & 1992 Apr 17  &    \nl
G275.25$+$0.0   &  09 35 25.5  &  $-$52 00 37   &  1990 Feb 12  &    \phz &     G321.375$-$0.9 &  15 19 45.9 & $-$58 21 26 & 1986 May 19  &    \nl
G275.625$-$0.9 &  09 33 15.4  &  $-$52 55 39   &  1990 Feb 13  &    \phz &      G321.375$+$0.9  &  15 12 42.2 & $-$56 49 33 & 1984 Apr 09  &    \nl
G275.625$+$0.9  &  09 41 06.3  &  $-$51 35 12   &  1986 Mar 04  &    \phz &     G321.4$-$0.5   &  15 18 24.4 & $-$58 00 56 & 1991 May 01  &  N \nl
G276.0$+$0.0    &  09 39 04.6  &  $-$52 30 38   &  1990 Feb 14  &    \phz &     G321.75$+$0.0   &  15 18 31.7 & $-$57 23 46 & 1992 Apr 03  &    \nl
G276.375$-$0.9 &  09 36 57.6  &  $-$53 25 55   &  1990 Feb 16  &    \phz &      G322.125$-$0.9 &  15 24 32.1 & $-$57 56 57 & 1987 May 07  &    \nl
G276.375$+$0.9  &  09 44 47.1  &  $-$52 04 33   &  1990 Feb 17  &    \phz &     G322.125$+$0.9  &  15 17 21.3 & $-$56 26 04 & 1984 Apr 11  &    \nl
G276.75$+$0.0   &  09 42 48.7  &  $-$53 00 13   &  1990 Feb 15  &    \phz &     G322.2$+$0.6    &  15 18 40.4 & $-$56 38 55 & 1994 Apr 19  &  N \nl
G277.125$-$0.9 &  09 40 45.1  &  $-$53 55 45   &  1990 Feb 18  &    \phz &      G322.5$+$0.0    &  15 23 11.3 & $-$56 59 28 & 1992 Apr 07  &    \nl
G277.125$+$0.9  &  09 48 32.7  &  $-$52 33 27   &  1990 Feb 19  &    \phz &     G322.7$-$0.2   &  15 25 23.4 & $-$57 00 33 & 1991 Apr 18  &  N \nl
G277.5$+$0.0    &  09 46 38.0  &  $-$53 29 21   &  1990 Feb 20  &    \phz &     G322.875$-$0.9 &  15 29 11.8 & $-$57 31 49 & 1987 May 08  &    \nl
G277.875$-$0.9 &  09 44 38.0  &  $-$54 25 08   &  1986 Mar 05  &    \phz &      G322.875$+$0.9  &  15 21 54.6 & $-$56 01 56 & 1984 Apr 10  &    \nl
G277.875$+$0.9  &  09 52 23.4  &  $-$53 01 54   &  1986 Mar 10  &    \phz &     G323.25$+$0.0   &  15 27 44.9 & $-$56 34 32 & 1986 May 22  &    \nl
G278.25$+$0.0   &  09 50 32.5  &  $-$53 58 01   &  1990 Feb 22  &    \phz &     G323.625$-$0.9 &  15 33 45.1 & $-$57 06 04 & 1987 May 14  &    \nl
G278.625$-$0.9 &  09 48 36.5  &  $-$54 54 02   &  1990 Feb 23  &    \phz &      G323.625$+$0.9  &  15 26 22.2 & $-$55 37 11 & 1987 May 25  &    \nl
G278.625$+$0.9  &  09 56 19.2  &  $-$53 29 51   &  1990 Feb 26  &    \phz &     G324.0$+$0.0    &  15 32 12.4 & $-$56 08 59 & 1992 Apr 08  &    \nl
G279.0$+$0.0    &  09 54 32.4  &  $-$54 26 12   &  1990 Feb 27  &    \phz &     G324.375$-$0.9 &  15 38 12.2 & $-$56 39 44 & 1987 May 18  &    \nl
              &              &               &  1991 Mar 04  &    \phz &        G324.375$+$0.9  &  15 30 44.1 & $-$55 11 49 & 1987 May 26  &    \nl
G279.375$-$0.9 &  09 52 40.6  &  $-$55 22 27   &  1991 Mar 28  &    \phz &      G324.75$+$0.0   &  15 36 34.0 & $-$55 42 51 & 1987 May 28  &    \nl
G279.375$+$0.9  &  10 00 20.2  &  $-$53 57 19   &  1991 Mar 08  &    \phz &     G325.125$-$0.9 &  15 42 33.0 & $-$56 12 48 & 1987 May 27  &    \nl
G279.75$+$0.0   &  09 58 37.8  &  $-$54 53 53   &  1991 Mar 07  &    \phz &     G325.125$+$0.9  &  15 35 00.4 & $-$54 45 53 & 1988 May 07  &    \nl
G280.125$-$0.9 &  09 56 50.5  &  $-$55 50 22   &  1991 Mar 05  &    \phz &      G325.4$+$0.9    &  15 36 49.5 & $-$54 39 53 & 1991 May 02  &  N \nl
G280.125$+$0.9  &  10 04 26.5  &  $-$54 24 15   &  1992 Feb 17  &    \phz &     G325.5$+$0.0    &  15 40 49.7 & $-$55 16 09 & 1987 May 29  &    \nl
G280.5$+$0.0    &  10 02 48.9  &  $-$55 21 02   &  1989 Feb 04  &    \phz &     G325.875$-$0.9 &  15 46 47.8 & $-$55 45 20 & 1990 May 19  &    \nl
G280.875$-$0.9 &  10 01 06.4  &  $-$56 17 44   &  1993 Mar 28  &    \phz &      G325.875$+$0.9  &  15 39 11.2 & $-$54 19 24 & 1990 May 16  &    \nl
G280.875$+$0.9  &  10 08 38.3  &  $-$54 50 38   &  1988 Mar 30  &    \phz &     G326.25$+$0.0   &  15 44 59.7 & $-$54 48 54 & 1990 May 18  &    \nl
              &              &               &  1989 Feb 06  &    \phz &        G326.3$-$1.6   &  15 52 54.8 & $-$56 13 25 & 1992 Apr 09  &  N \nl
G281.25$+$0.0   &  10 07 05.7  &  $-$55 47 38   &  1986 Mar 11  &    \phz &     G326.625$-$0.9 &  15 50 56.7 & $-$55 17 20 & 1991 Apr 23  &    \nl
G281.625$-$0.9 &  10 05 28.4  &  $-$56 44 33   &  1985 Feb 04  &    \phz &      G326.625$+$0.9  &  15 43 16.5 & $-$53 52 22 & 1990 May 17  &    \nl
G281.625$+$0.9  &  10 12 55.6  &  $-$55 16 28   &  1988 Mar 31  &    \phz &                   &             &            & 1992 Jul 08  &    \nl
G282.0$-$1.8   &  10 03 43.4  &  $-$57 39 26   &  1991 Mar 11  &  N \phz &      G327.0$+$0.0    &  15 49 04.1 & $-$54 21 08 & 1991 Apr 22  &    \nl
G282.0$-$1.2   &  10 06 44.3  &  $-$57 12 41   &  1994 Mar 01  &  N \phz &      G327.375$-$0.9 &  15 54 59.8 & $-$54 48 50 & 1992 Apr 10  &    \nl
G282.0$+$0.0    &  10 11 28.3  &  $-$56 13 40   &  1986 Mar 12  &    \phz &     G327.375$+$0.9  &  15 47 16.6 & $-$53 24 49 & 1990 May 20  &    \nl
G282.375$-$0.9 &  10 09 56.6  &  $-$57 10 48   &  1985 Feb 14  &    \phz &      G327.75$+$0.0   &  15 53 03.1 & $-$53 52 51 & 1993 Apr 20  &    \nl
G282.375$+$0.9  &  10 17 18.5  &  $-$55 41 43   &  1989 Feb 08  &    \phz &     G328.125$-$0.9 &  15 58 57.2 & $-$54 19 50 & 1991 Apr 27  &    \nl
G282.75$+$0.0   &  10 15 56.9  &  $-$56 39 06   &  1988 Mar 14  &    \phz &     G328.125$+$0.9  &  15 51 11.4 & $-$52 56 46 & 1991 Apr 25  &    \nl
G283.125$-$0.9 &  10 14 31.0  &  $-$57 36 26   &  1985 Feb 19  &    \phz &      G328.5$+$0.0    &  15 56 56.6 & $-$53 24 06 & 1993 Apr 21  &    \nl
G283.125$+$0.9  &  10 21 47.1  &  $-$56 06 22   &  1987 Mar 10  &    \phz &     G328.6$-$0.4   &  15 58 51.6 & $-$53 38 33 & 1991 Jun 01  &  N \nl
G283.5$+$0.0    &  10 20 31.5  &  $-$57 03 56   &  1993 Mar 01  &    \phz &     G328.875$-$0.9 &  16 02 49.1 & $-$53 50 22 & 1989 May 04  &    \nl
G283.875$-$0.9 &  10 19 11.9  &  $-$58 01 27   &  1992 Mar 03  &    \phz &      G328.875$+$0.9  &  15 55 01.2 & $-$52 28 15 & 1991 Apr 26  &    \nl
G283.875$+$0.9  &  10 26 21.5  &  $-$56 30 22   &  1987 Mar 09  &    \phz &     G329.25$+$0.0   &  16 00 44.9 & $-$52 54 53 & 1992 May 04  &    \nl
G284.0$+$0.8    &  10 26 33.3  &  $-$56 40 19   &  1991 Mar 13  &  N \phz &     G329.625$-$0.9 &  16 06 35.6 & $-$53 20 27 & 1989 May 10  &    \nl
G284.25$+$0.0   &  10 25 12.3  &  $-$57 28 07   &  1983 Mar 10  &    \phz &     G329.625$+$0.9  &  15 58 45.9 & $-$51 59 15 & 1989 May 09  &    \nl
G284.3$-$0.4   &  10 24 20.5  &  $-$57 49 15   &  1994 Mar 02  &  N \phz &      G329.9$+$0.5    &  16 01 48.0 & $-$52 08 22 & 1990 Apr 26  &  N \nl
G284.4$-$0.5   &  10 24 35.6  &  $-$57 56 43   &  1987 Apr 06  &  N \phz &      G330.0$+$0.0    &  16 04 28.1 & $-$52 25 13 & 1992 Apr 12  &    \nl
              &              &               &  1987 Apr 14  &  N \phz &                      &             &            & 1993 Apr 22  &    \nl
G284.625$-$0.9 &  10 23 59.3  &  $-$58 25 48   &  1985 Feb 25  &    \phz &      G330.2$-$0.3   &  16 06 49.8 & $-$52 28 03 & 1991 May 07  &  N \nl
G285.625$+$0.0  &  10 27 35.1  &  $-$57 40 00   &  1984 Mar 31  &  N \phz &     G330.375$-$0.9 &  16 10 16.9 & $-$52 50 07 & 1988 May 10  &    \nl
G284.625$+$0.9  &  10 31 01.7  &  $-$56 53 44   &  1993 Mar 03  &    \phz &     G330.375$+$0.9  &  16 02 25.8 & $-$51 29 50 & 1988 May 05  &    \nl
G285.0$+$0.0    &  10 29 59.2  &  $-$57 51 38   &  1992 Mar 04  &    \phz &     G330.75$+$0.0   &  16 08 06.3 & $-$51 55 07 & 1984 May 08  &    \nl
G285.37$+$0.1   &  10 32 53.9  &  $-$57 55 29   &  1992 Mar 05  &  N \phz &     G331.125$-$0.9 &  16 13 53.1 & $-$52 19 22 & 1991 May 08  &    \nl
G285.375$-$0.9 &  10 28 53.3  &  $-$58 49 29   &  1993 Mar 02  &    \phz &      G331.125$+$0.9  &  16 06 00.9 & $-$50 59 58 & 1988 May 08  &    \nl
G285.375$+$0.9  &  10 35 47.7  &  $-$57 16 26   &  1993 Mar 05  &    \phz &     G331.5$+$0.0    &  16 11 39.7 & $-$51 24 37 & 1992 Jul 25  &    \nl
G285.4$+$0.0    &  10 32 53.7  &  $-$58 05 29   &  1991 Mar 15  &  N \phz &                   &             &            & 1993 Apr 25  &    \nl
              &              &               &  1991 Apr 02  &  N \phz &        G331.875$-$0.9 &  16 17 24.4 & $-$51 48 13 & 1991 May 09  &    \nl
G285.5$-$1.0   &  10 29 20.8  &  $-$58 55 24   &  1991 Mar 14  &  N \phz &      G331.875$+$0.9  &  16 09 31.4 & $-$50 29 43 & 1993 May 02  &    \nl
G285.75$+$0.0   &  10 34 52.3  &  $-$58 14 28   &  1993 Mar 04  &    \phz &     G332.25$+$0.0   &  16 15 08.3 & $-$50 53 43 & 1992 Jul 31  &    \nl
G286.125$-$0.9 &  10 33 53.9  &  $-$59 12 28   &  1988 Mar 21  &    \phz &                    &             &            & 1993 Apr 27  &    \nl
G286.125$+$0.9  &  10 40 39.6  &  $-$57 38 25   &  1989 Feb 11  &    \phz &     G332.625$-$0.9 &  16 20 50.9 & $-$51 16 41 & 1993 May 05  &    \nl
G286.29$-$0.8  &  10 35 33.1  &  $-$50 10 34   &  1992 Feb 23  &  N \phz &      G332.625$+$0.9  &  16 12 57.4 & $-$49 59 03 & 1993 May 03  &    \nl
G286.5$-$2.4   &  10 30 38.4  &  $-$60 39 26   &  1990 Mar 01  &  N \phz &      G333.0$+$0.0    &  16 18 32.3 & $-$50 22 26 & 1993 May 04  &    \nl
G286.5$+$0.0    &  10 39 51.7  &  $-$58 36 34   &  1985 Feb 27  &    \phz &     G333.375$-$0.9 &  16 24 12.7 & $-$50 44 48 & 1993 May 06  &    \nl
G286.6$+$0.0    &  10 40 32.3  &  $-$58 36 35   &  1990 Apr 07  &  N \phz &     G333.375$+$0.9  &  16 16 18.9 & $-$49 28 01 & 1988 May 28  &    \nl
G286.7$-$1.2   &  10 36 12.5  &  $-$59 45 35   &  1990 Mar 02  &  N \phz &      G333.5$-$0.5   &  16 22 46.9 & $-$50 22 00 & 1991 Apr 29  &  N \nl
G286.875$-$0.9 &  10 39 01.1  &  $-$59 34 43   &  1985 Mar 01  &    \phz &      G333.75$+$0.0   &  16 21 51.9 & $-$49 50 48 & 1992 Apr 15  &    \nl
G286.875$+$0.9  &  10 45 37.4  &  $-$57 59 41   &  1989 Feb 12  &    \phz &     G334.125$-$0.9 &  16 27 30.0 & $-$50 12 34 & 1993 May 31  &    \nl
G287.17$-$1.6  &  10 38 22.5  &  $-$60 19 38   &  1992 Feb 24  &  N \phz &      G334.125$+$0.9  &  16 19 36.2 & $-$48 56 37 & 1988 May 29  &    \nl
G287.25$+$0.0   &  10 44 57.4  &  $-$58 57 57   &  1985 Mar 05  &    \phz &     G334.4$+$0.7    &  16 21 43.2 & $-$48 57 04 & 1991 May 31  &  N \nl
G287.5$+$0.6    &  10 48 29.9  &  $-$58 32 53   &  1990 Mar 07  &  N \phz &     G334.5$+$0.0    &  16 25 07.1 & $-$49 18 48 & 1992 Jul 16  &    \nl
G287.6$-$2.1   &  10 39 11.9  &  $-$60 59 39   &  1990 Mar 03  &  N \phz &                    &             &            & 1993 May 10  &    \nl
G287.625$-$0.9 &  10 44 15.1  &  $-$59 56 12   &  1992 Feb 19  &    \phz &      G334.875$-$0.9 &  16 30 42.9 & $-$49 39 59 & 1988 Jun 03  &    \nl
              &              &               &  1993 Mar 08  &    \phz &        G334.875$+$0.9  &  16 22 49.4 & $-$48 24 52 & 1991 May 13  &    \nl
G287.625$+$0.9  &  10 50 41.0  &  $-$58 20 12   &  1989 Feb 13  &    \phz &     G335.25$+$0.0   &  16 28 18.2 & $-$48 46 29 & 1989 May 12  &    \nl
G288.0$+$0.0    &  10 50 09.3  &  $-$59 18 34   &  1993 Mar 09  &    \phz &     G335.75$-$0.9  &  16 34 22.7 & $-$49 01 35 & 1989 May 15  &    \nl
G288.35$-$0.9  &  10 49 27.9  &  $-$60 15 54   &  1992 Feb 25  &  N \phz &      G335.75$+$0.9   &  16 26 29.6 & $-$47 47 24 & 1989 May 11  &    \nl
G288.375$-$0.9 &  10 49 35.7  &  $-$60 16 54   &  1992 Feb 26  &    \phz &      G336.0$-$0.8   &  16 34 44.5 & $-$48 46 12 & 1991 Apr 30  &  N \nl
G288.375$+$0.9  &  10 55 50.6  &  $-$58 39 57   &  1989 Feb 14  &    \phz &     G336.25$+$0.0   &  16 32 26.6 & $-$48 02 53 & 1993 May 11  &    \nl
G288.5$-$1.6   &  10 47 46.1  &  $-$60 59 52   &  1990 Mar 06  &  N \phz &      G336.5$-$1.3   &  16 39 14.3 & $-$48 45 49 & 1992 May 06  &  N \nl
G288.5$-$0.2   &  10 53 10.5  &  $-$59 45 59   &  1990 Mar 08  &  N \phz &      G336.75$-$0.9  &  16 38 27.0 & $-$48 17 11 & 1993 Apr 28  &    \nl
G288.75$+$0.0   &  10 55 27.5  &  $-$59 38 23   &  1985 Mar 13  &    \phz &     G336.75$+$0.9   &  16 30 34.8 & $-$47 04 03 & 1993 May 27  &    \nl
G289.125$-$0.9 &  10 55 03.0  &  $-$60 36 48   &  1989 Feb 26  &    \phz &      G337.25$+$0.0   &  16 36 28.0 & $-$47 18 44 & 1990 Jun 08  &    \nl
G289.125$+$0.9  &  11 01 05.9  &  $-$58 58 54   &  1989 Feb 17  &    \phz &     G337.5$+$1.6    &  16 30 07.5 & $-$46 01 30 & 1992 May 05  &  N \nl
G289.5$+$0.0    &  11 00 51.9  &  $-$59 57 23   &  1989 Feb 16  &    \phz &     G337.75$-$0.9  &  16 42 24.4 & $-$47 32 17 & 1993 Jun 01  &    \nl
G289.875$-$0.9 &  11 00 37.0  &  $-$60 55 51   &  1983 Mar 14  &    \phz &      G337.75$+$0.9   &  16 34 33.4 & $-$46 20 09 & 1989 May 18  &    \nl
              &              &               &  1989 Feb 18  &    \phz &        G338.1$+$0.1    &  16 39 19.7 & $-$46 32 53 & 1990 Jun 09  &  N \nl
G289.875$+$0.9  &  11 06 27.0  &  $-$59 17 02   &  1989 Feb 19  &    \phz &     G338.25$+$0.0   &  16 40 22.8 & $-$46 34 05 & 1992 Apr 16  &    \nl
G289.9$+$0.0    &  11 03 35.0  &  $-$60 06 11   &  1991 Mar 22  &  N \phz &     G338.75$-$0.9  &  16 46 15.1 & $-$46 46 53 & 1989 May 23  &    \nl
G290.0$-$1.1   &  11 00 37.3  &  $-$61 10 08   &  1991 Mar 19  &  N \phz &      G338.75$+$0.9   &  16 38 25.5 & $-$45 35 45 & 1993 May 12  &    \nl
G290.25$+$0.0   &  11 06 22.4  &  $-$60 15 32   &  1989 Feb 20  &    \phz &     G339.2$-$0.6   &  16 46 39.8 & $-$46 15 23 & 1991 May 16  &  N \nl
G290.5$+$0.1    &  11 08 27.3  &  $-$60 16 16   &  1991 Mar 25  &  N \phz &     G339.2$+$0.5    &  16 41 37.6 & $-$45 30 44 & 1991 Jun 02  &  N \nl
G290.625$-$0.9 &  11 06 17.5  &  $-$61 14 02   &  1992 Jun 13  &    \phz &      G339.25$+$0.0   &  16 44 11.2 & $-$45 48 56 & 1992 Apr 20  &    \nl
G290.625$+$0.9  &  11 11 53.7  &  $-$59 34 19   &  1984 Mar 06  &    \phz &     G339.75$-$0.9  &  16 49 59.5 & $-$46 01 02 & 1990 Jun 11  &    \nl
G291.0$+$0.0    &  11 11 58.9  &  $-$60 32 49   &  1992 Jun 05  &    \phz &     G339.75$+$0.9   &  16 42 11.5 & $-$44 50 51 & 1989 May 26  &    \nl
G291.375$-$0.9 &  11 12 04.4  &  $-$61 31 19   &  1991 Oct 05  &    \phz &      G340.25$+$0.0   &  16 47 53.5 & $-$45 03 20 & 1986 Jun 26  &    \nl
G291.375$+$0.9  &  11 17 25.9  &  $-$59 50 44   &  1989 Feb 21  &    \phz &     G340.75$-$0.9  &  16 53 37.8 & $-$45 14 44 & 1990 Jun 12  &    \nl
G291.4$-$0.6   &  11 13 38.7  &  $-$61 16 21   &  1994 Mar 03  &  N \phz &      G340.75$+$0.9   &  16 45 51.6 & $-$44 05 30 & 1990 Jun 10  &    \nl
G291.6$-$0.5   &  11 15 09.6  &  $-$61 16 22   &  1991 Aug 16  &  N \phz &      G341.25$+$0.0   &  16 51 29.9 & $-$44 17 17 & 1990 May 30  &    \nl
G291.75$+$0.0   &  11 17 41.2  &  $-$60 49 12   &  1993 Mar 08  &    \phz &                   &             &            & 1990 Jul 01  &    \nl
G292.125$-$0.9 &  11 17 57.5  &  $-$61 47 40   &  1992 Jun 28  &    \phz &      G341.75$-$0.9  &  16 57 10.3 & $-$44 28 02 & 1990 Jul 13  &    \nl
G292.125$+$0.9  &  11 23 03.5  &  $-$60 06 15   &  1989 Feb 24  &    \phz &     G341.75$+$0.9   &  16 49 26.1 & $-$43 19 42 & 1990 May 31  &    \nl
G292.38$+$0.0   &  11 22 28.8  &  $-$61 04 39   &  1990 Mar 10  &  N \phz &     G342.25$+$0.0   &  16 55 00.7 & $-$43 30 50 & 1990 Jul 27  &    \nl
              &              &               &  1992 Mar 06  &  N \phz &        G342.75$-$0.9  &  17 00 37.3 & $-$43 40 57 & 1989 Jun 01  &    \nl
G292.5$+$0.0    &  11 23 29.4  &  $-$61 04 40   &  1989 Feb 22  &    \phz &     G342.75$+$0.9   &  16 52 55.1 & $-$42 33 29 & 1990 Jul 20  &    \nl
G292.875$-$0.9 &  11 23 56.8  &  $-$62 03 05   &  1984 Mar 11  &    \phz &      G343.1$-$0.6   &  17 00 36.2 & $-$43 14 24 & 1990 Jun 13  &  N \nl
G292.875$+$0.9  &  11 28 46.4  &  $-$60 20 51   &  1992 Apr 01  &    \phz &     G343.25$+$0.0   &  16 58 26.1 & $-$42 43 59 & 1990 Jul 10  &    \nl
G293.25$+$0.0   &  11 29 23.0  &  $-$61 19 11   &  1992 Mar 31  &    \phz &     G343.75$-$0.9  &  17 03 58.9 & $-$42 53 29 & 1989 Jun 02  &    \nl
G293.625$-$0.9 &  11 30 02.0  &  $-$62 17 30   &  1992 Mar 11  &    \phz &      G343.75$+$0.9   &  16 56 18.9 & $-$41 46 52 & 1989 May 31  &    \nl
G293.625$+$0.9  &  11 34 34.2  &  $-$60 34 31   &  1992 Jun 08  &    \phz &     G344.25$+$0.0   &  17 01 46.4 & $-$41 56 46 & 1990 Jul 11  &    \nl
G293.83$-$0.4  &  11 33 18.6  &  $-$61 46 35   &  1992 Mar 10  &  N \phz &      G344.4$-$0.3   &  17 03 32.1 & $-$42 04 12 & 1991 May 17  &  N \nl
G294.0$+$0.0    &  11 35 22.0  &  $-$61 32 44   &  1989 Mar 03  &    \phz &     G344.75$-$0.9  &  17 07 15.6 & $-$42 05 41 & 1989 Jun 04  &    \nl
G294.2$-$0.1   &  11 36 22.4  &  $-$61 42 44   &  1990 Mar 11  &  N \phz &      G344.75$+$0.9   &  16 59 37.8 & $-$40 59 53 & 1989 May 30  &    \nl
G294.375$-$0.9 &  11 36 12.9  &  $-$62 30 55   &  1989 Mar 22  &    \phz &      G345.25$+$0.0   &  17 05 01.8 & $-$41 09 11 & 1992 Jun 01  &    \nl
G294.375$+$0.9  &  11 40 26.8  &  $-$60 47 13   &  1989 Mar 07  &    \phz &     G345.4$-$0.9   &  17 09 34.4 & $-$41 35 46 & 1994 Apr 21  &  N \nl
G294.75$+$0.0   &  11 41 26.0  &  $-$61 45 17   &  1989 Mar 08  &    \phz &     G345.75$-$0.9  &  17 10 27.3 & $-$41 17 33 & 1990 Jul 12  &    \nl
G295.125$-$0.9 &  11 42 29.1  &  $-$62 43 19   &  1992 Mar 12  &    \phz &      G345.75$+$0.9   &  17 02 51.9 & $-$40 12 32 & 1992 May 07  &    \nl
G295.125$+$0.9  &  11 46 24.0  &  $-$60 58 56   &  1989 Mar 09  &    \phz &     G346.0$+$0.5    &  17 05 28.6 & $-$40 14 04 & 1991 May 21  &  N \nl
G295.5$+$0.0    &  11 47 34.8  &  $-$61 56 49   &  1989 Mar 10  &    \phz &     G346.25$+$0.0   &  17 08 12.6 & $-$40 21 17 & 1993 May 13  &    \nl
G295.875$-$0.9 &  11 48 50.3  &  $-$62 54 39   &  1990 Mar 12  &    \phz &      G346.75$-$0.9  &  17 13 34.5 & $-$40 29 05 & 1993 Jun 02  &    \nl
G295.875$+$0.9  &  11 52 25.4  &  $-$61 09 38   &  1989 Mar 13  &    \phz &     G346.75$+$0.9   &  17 06 01.5 & $-$39 24 51 & 1991 May 22  &    \nl
G296.25$+$0.0   &  11 53 48.1  &  $-$62 07 18   &  1989 Mar 14  &    \phz &     G347.25$+$0.0   &  17 11 18.8 & $-$39 33 03 & 1990 Jul 15  &    \nl
G296.625$-$0.9 &  11 55 16.3  &  $-$63 04 55   &  1989 Mar 15  &    \phz &      G347.75$-$0.9  &  17 16 37.3 & $-$39 40 20 & 1992 Apr 30  &    \nl
G296.625$+$0.9  &  11 58 30.7  &  $-$61 19 19   &  1989 Mar 16  &    \phz &     G347.75$+$0.9   &  17 09 06.7 & $-$38 36 51 & 1991 Jun 07  &    \nl
G297.0$+$0.0    &  12 00 05.5  &  $-$62 16 44   &  1989 Mar 20  &    \phz &     G348.25$+$0.0   &  17 14 20.8 & $-$38 44 31 & 1992 Jul 21  &    \nl
G297.2$+$0.3    &  12 02 34.5  &  $-$61 59 43   &  1991 Jul 27  &  N \phz &     G348.75$-$0.9  &  17 19 35.9 & $-$38 51 18 & 1992 May 13  &    \nl
G297.375$-$0.9 &  12 01 46.5  &  $-$63 14 05   &  1992 Mar 17  &    \phz &      G348.75$+$0.9   &  17 12 07.8 & $-$37 48 32 & 1991 Jun 10  &    \nl
G297.375$+$0.9  &  12 04 39.6  &  $-$61 27 58   &  1989 Mar 21  &    \phz &     G349.25$+$0.0   &  17 17 18.6 & $-$37 55 42 & 1990 Jul 16  &    \nl
G297.6$-$0.9   &  12 03 30.0  &  $-$63 14 43   &  1991 Jul 19  &  N \phz &      G349.75$-$0.9  &  17 22 30.5 & $-$38 02 00 & 1991 May 27  &    \nl
G297.75$+$0.0   &  12 06 26.6  &  $-$62 25 06   &  1989 Mar 23  &    \phz &     G349.75$+$0.9   &  17 15 04.8 & $-$36 59 55 & 1992 May 12  &    \nl
G298.125$-$0.9 &  12 08 20.7  &  $-$63 22 08   &  1992 Mar 20  &    \phz &      G349.9$-$2.3   &  17 28 56.6 & $-$38 42 23 & 1994 Jul 05  &  N \nl
G298.125$+$0.9  &  12 10 51.7  &  $-$61 35 33   &  1989 Apr 03  &    \phz &     G350.25$+$0.0   &  17 20 12.6 & $-$37 06 36 & 1991 May 12  &    \nl
G298.5$-$1.3   &  12 12 40.2  &  $-$62 52 41   &  1994 Mar 10  &  N \phz &                    &             &            & 1992 Sep 13  &    \nl
G298.5$+$0.0    &  12 12 51.1  &  $-$62 32 21   &  1989 Apr 01  &    \phz &     G350.3$-$0.7   &  17 23 24.2 & $-$37 27 47 & 1991 May 28  &  N \nl
G298.875$-$0.9 &  12 14 58.2  &  $-$63 29 03   &  1992 Jun 27  &    \phz &      G350.4$-$0.8   &  17 23 54.3 & $-$37 27 45 & 1991 May 29  &  N \nl
G298.875$+$0.9  &  12 17 06.7  &  $-$61 42 04   &  1989 Apr 02  &    \phz &     G350.75$-$0.9  &  17 25 21.3 & $-$37 12 27 & 1991 Jun 17  &    \nl
G299.25$+$0.0   &  12 19 18.4  &  $-$62 38 31   &  1993 Mar 11  &    \phz &     G350.75$+$0.9   &  17 17 58.1 & $-$36 11 02 & 1991 May 23  &    \nl
G299.625$-$0.9 &  12 21 38.7  &  $-$63 34 49   &  1984 Mar 21  &    \phz &                    &             &            & 1993 May 18  &    \nl
G299.625$+$0.9  &  12 23 24.2  &  $-$61 47 30   &  1989 Apr 04  &    \phz &     G351.25$+$0.0   &  17 23 02.9 & $-$36 17 15 & 1991 Jun 16  &    \nl
G300.0$+$0.0    &  12 25 48.2  &  $-$62 43 33   &  1989 Apr 07  &    \phz &     G351.4$-$0.4   &  17 25 22.4 & $-$36 22 38 & 1991 Jun 21  &  N \nl
              &              &               &  1993 Mar 12  &    \phz &        G351.75$-$0.9  &  17 28 08.4 & $-$36 22 39 & 1991 May 19  &    \nl
G300.375$-$0.9 &  12 28 21.6  &  $-$63 39 26   &  1989 Apr 09  &    \phz &      G351.75$+$0.9   &  17 20 47.7 & $-$35 21 53 & 1983 Jul 04  &    \nl
G300.375$+$0.9  &  12 29 43.6  &  $-$61 51 50   &  1989 Apr 08  &    \phz &                   &             &            & 1992 May 14  &    \nl
G300.5$-$0.3   &  12 29 50.0  &  $-$63 06 35   &  1991 Jul 05  &  N \phz &      G352.25$+$0.0   &  17 25 49.5 & $-$35 27 39 & 1992 May 18  &    \nl
G300.75$+$0.0   &  12 32 20.0  &  $-$62 47 27   &  1984 Mar 22  &    \phz &     G352.75$-$0.9  &  17 30 52.1 & $-$35 32 38 & 1992 May 21  &    \nl
G300.87$+$0.9   &  12 33 56.4  &  $-$61 54 07   &  1983 Apr 14  &  N \phz &     G352.75$+$0.9   &  17 23 33.8 & $-$34 32 29 & 1988 Jun 08  &    \nl
G301.125$-$0.9 &  12 35 06.5  &  $-$63 42 52   &  1988 Apr 06  &    \phz &      G353.25$+$0.0   &  17 28 32.8 & $-$34 37 49 & 1992 May 31  &    \nl
G301.125$+$0.9  &  12 36 04.6  &  $-$61 55 04   &  1988 Apr 05  &    \phz &     G353.4$-$0.3   &  17 30 19.6 & $-$34 42 17 & 1991 May 30  &  N \nl
G301.5$-$1.0   &  12 38 15.4  &  $-$63 49 30   &  1991 Jun 14  &  N \phz &      G353.75$-$0.9  &  17 33 32.4 & $-$34 42 24 & 1993 May 19  &    \nl
G301.5$+$0.0    &  12 38 53.2  &  $-$62 50 13   &  1992 Mar 18  &    \phz &     G353.75$+$0.9   &  17 26 16.6 & $-$33 42 51 & 1987 Jun 21  &    \nl
G301.875$-$0.9 &  12 41 52.7 & $-$63 45 08 & 1990 Mar 13  &    \phz &   G354.25$+$0.0   &  17 31 12.9 & $-$33 47 46 & 1985 Jun 11  &    \nl
G301.875$+$0.9  &  12 42 26.8 & $-$61 57 12 & 1993 Mar 16  &    \phz &                &             &            & 1992 May 22  &    \nl
G302.1$+$0.9    &  12 44 26.9 & $-$61 56 25 & 1990 Apr 01  &  N \phz &  G354.75$-$0.9  &  17 36 09.6 & $-$33 51 57 & 1987 Jun 22  &    \nl
G302.25$+$0.0   &  12 45 27.4 & $-$62 51 50 & 1983 Apr 05  &    \phz &  G354.75$+$0.9   &  17 28 56.2 & $-$32 52 59 & 1993 May 26  &    \nl
G302.625$-$0.9 &  12 48 39.6 & $-$63 46 12 & 1985 Apr 09  &    \phz &   G354.8$-$0.2   &  17 33 24.5 & $-$33 23 21 & 1987 Jul 01  &  N \nl
G302.625$+$0.9  &  12 48 49.6 & $-$61 58 13 & 1986 Apr 08  &    \phz &  G355.25$+$0.0   &  17 33 49.8 & $-$32 57 30 & 1992 May 28  &    \nl
              &             &            & 1989 Apr 29  &    \phz &     G355.75$-$0.9  &  17 38 43.8 & $-$33 01 19 & 1991 Aug 09  &    \nl
G303.0$+$0.0    &  12 52 02.1 & $-$62 52 18 & 1988 Apr 09  &    \phz &  G355.75$+$0.9   &  17 31 32.8 & $-$32 02 54 & 1987 Jun 28  &    \nl
G303.375$-$0.9 &  12 55 26.8 & $-$63 46 06 & 1985 Apr 10  &    \phz &   G356.25$+$0.0   &  17 36 23.8 & $-$32 07 02 & 1985 Jun 13  &    \nl
G303.375$+$0.9  &  12 55 12.5 & $-$61 58 07 & 1991 Apr 07  &    \phz &  

\enddata
\end{deluxetable}